\documentclass[modern]{aastex631}

\begin{document}

\title{Exploring the Extremes: Characterizing a New Population of Old and Cold Brown Dwarfs}

\author[0000-0002-1125-7384]{Aaron M. Meisner}
\affiliation{NSF's National Optical-Infrared Astronomy Research Laboratory, 950 N. Cherry Avenue, Tucson, AZ 85719, USA}

\author[0000-0002-3681-2989]{S. K. Leggett}
\affiliation{Gemini Observatory/NSF's NOIRLab, 670 N. A'ohoku Place, Hilo, HI 96720, USA}

\author[0000-0002-9632-9382]{Sarah E. Logsdon}
\affiliation{NSF's National Optical-Infrared Astronomy Research Laboratory, 950 N. Cherry Avenue, Tucson, AZ 85719, USA}

\author[0000-0002-6294-5937]{Adam C. Schneider}
\affiliation{US Naval Observatory, Flagstaff Station, P.O. Box 1149, Flagstaff, AZ 86002, USA}

\author[0000-0001-6172-3403]{Pascal Tremblin}
\affiliation{Universite Paris-Saclay, UVSQ, CNRS, CEA, Maison de la Simulation, 91191, Gif-sur-Yvette, France}

\author[0000-0001-6041-7092]{Mark Phillips}
\affiliation{Institute for Astronomy, University of Hawaii, 2680 Woodlawn Drive, Honolulu, HI 96822, USA}

\begin{abstract}

Mapping out the populations of thick disk and halo brown dwarfs is important for understanding the metallicity dependence of low-temperature atmospheres and the substellar mass function. Recently, a new population of cold and metal-poor brown dwarfs has been discovered, with $T_{\rm{eff}}$ $\lesssim$ 1400 K and metallicity $\lesssim$ $-$1 dex. This population includes what may be the first known ``extreme T-type subdwarfs'' and possibly the first Y-type subdwarf, WISEA J153429.75$-$104303.3. We have conducted a Gemini YJHK/Ks photometric follow-up campaign targeting potentially metal-poor T and Y dwarfs, utilizing the GNIRS and Flamingos-2 instruments. We present 14 near-infrared photometric detections of 8 unique targets: six T subdwarf candidates, one moderately metal poor Y dwarf candidate, and one Y subdwarf candidate. We have obtained the first ever ground-based detection of the highly anomalous object WISEA J153429.75$-$104303.3. The F110W$-$$J$ color of WISEA J153429.75$-$104303.3 is significantly bluer than that of other late-T and Y dwarfs, indicating that WISEA J153429.75$-$104303.3 has an unusual spectrum in the 0.9-1.4~$\mu$m wavelength range which encompasses the $J$-band peak. Our $J$-band detection of WISEA J153429.75$-$104303.3 and corresponding model comparisons suggest a subsolar metallicity and temperature of 400-550 K for this object. JWST spectroscopic follow-up at near-infrared and mid-infrared wavelengths would allow us to better understand the spectral peculiarities of WISEA J153429.75$-$104303.3, assess its physical properties, and conclusively determine whether or not it is the first Y-type subdwarf.

\end{abstract}

\section{Introduction} \label{sec:intro}

In the solar neighborhood, and in young clusters, the mass function shows a continuum from dwarf stars with masses of tenths of a solar mass ($M_{\odot}$) to brown dwarfs (objects with insufficient mass for hydrogen burning) with masses of a few Jupiter masses  \citep[$M_J$;e.g.,][]{Gagne_2017,Kirkpatrick_2019,Kirkpatrick_2021a,Lodieu_2021,Luhman_2020}. Distant  $\sim 0.1 M_{\odot}$ objects have been found via microlensing in the Galactic bulge \citep{Chung_2017}, and Gaia data \citep{gaia_mission} show that a very low mass (VLM) population with masses down to $0.2 M_{\odot}$ exists in all components of the Galaxy \citep{Hallakoun_2021}. Furthermore, in the VLM star  regime, the favored initial mass function increases toward lower masses \citep[e.g.,][]{Chabrier_2003,Kroupa_2001}.

VLM stars and brown dwarfs are intrinsically faint, due to their small sizes and low effective temperatures ($T_{\rm eff}$), hence studies of the coldest brown dwarfs are limited to the solar neighborhood. Such studies show that brown dwarfs with  $T_{\rm eff} \lesssim 1000$~K range in mass from 5 to 50 $M_J$, in metallicity ([m/H]) from $-0.5$ to $+0.3$~dex, and in age from 0.5 to 5~Gyr   \citep{Leggett_2021, Line_2017}. Not surprisingly, the metallicity and age range of the sample is typical of the thin disk \citep{Hallakoun_2021, Kilic_2019}. Despite their pervasiveness, many questions remain regarding the properties of the lowest luminosity objects in our Milky Way Galaxy. For example, (1) how has the birth rate of VLM stars and brown dwarfs evolved over time, from early periods of star formation to the present epoch, and (2) how do brown dwarf atmospheres and spectral energy distributions (SEDs) change with temperature, metallicity, and age?  Identifying and studying nearby brown dwarfs with extreme values of $T_{\rm eff}$, [m/H], age, and mass is key to answering these questions.

Old brown dwarfs have cooled to very low temperatures. Simulations of the solar neighborhood show that the number of brown dwarfs older than 8~Gyr only becomes significant at $T_{\rm eff} \lesssim 1000$~K \citep{Kirkpatrick_2021a}. Brown dwarfs which are members of the thick disk or halo are expected to have large motions of $\gtrsim 100$~km/s, and [m/H] $< -0.5$ dex
\citep{Faherty_2009, Hallakoun_2021, Kilic_2019}.
Very few such objects are known, with a prime example being Wolf 1130C, a T8 companion to a subdwarf star system \citep{Mace_2013b,Mace_2018}. 
Recently, long-baseline mid-infrared (IR) images
from the Wide-field IR Survey Explorer \citep[WISE;][]{Wright_2010} and its continuation NEOWISE \citep{Mainzer_2011,Mainzer_2014} have become available, allowing for all-sky searches
for objects with large motions  at wavelengths of  $3 \lesssim \lambda ~\mu$m $\lesssim 5$, where cold brown dwarf emission peaks. 
These data have enabled the identification of candidate old and cold brown dwarfs.

The new objects, discovered by 
\cite{Brooks_2022}, \cite{Meisner_2020a, Meisner_2020b}, \cite{Schneider_2020}, and the Backyard Worlds: Planet 9 citizen science project \citep{kuchner_2017}, have high velocities of $\sim$200~km/s\footnote{Not all old and metal-poor stars have high velocities \citep[e.g.,][]{Haywood_2013}. WISEA J181006.18$-$101000.5, one of the new metal-poor T dwarf discoveries, has UVW velocities ranging from $-$45 to $-$29 km~s$^{-1}$ \citep{Lodieu_2022}.}, and $J$ ($1.25~\mu$m), 3.6~$\mu$m, and 4.5~$\mu$m colors indicative of extremely low metallicity ([m/H] $\leq -1$).   For the first time, the small population of thick disk and halo brown dwarfs near the Sun \citep[$\approx 10\%$ of dwarfs;][]{Bensby_2014}
with unusually low [m/H] can be explored. Understanding the physical and kinematic properties of the substellar subdwarf population is an important initial step toward determining the low-mass star formation rate as a function of global Galactic evolution of the Milky Way, especially at early stages in its formation.

The coldest metal-poor brown dwarf candidates are, by selection, quite difficult to observe at near-infrared (NIR) wavelengths. The largest aperture, most IR-capable facilities are required in order to obtain ground-based detections. Until now, many members of the new metal-poor population have lacked even a single NIR ($0.9~\mu\textrm{m} \lesssim \lambda \lesssim 2.5~\mu$m) photometric detection, with few having even one NIR color available. To remedy this situation, we have conducted a NIR photometric follow-up campaign utilizing GNIRS at Gemini North and Flamingos-2 at Gemini South, to fill in the NIR SEDs of the new metal-poor brown dwarf population and prepare/prioritize for future spectroscopic follow-up. We have obtained 14 photometric detections in the YJHK/Ks bandpasses for a sample of 8 cold and (potentially) metal-poor targets. We use this new information in combination with low-temperature, low-metallicity atmospheric models \citep{Tremblin_2015} to elucidate the physical parameters of these objects.

In $\S$\ref{sec:sample} we explain our selection of a cold, metal-poor brown dwarf candidate sample to follow up with Gemini imaging. In $\S$\ref{sec:observations} we detail our Gemini observations/reductions and present the resulting photometric measurements. In $\S$\ref{sec:discussion} we discuss the implications of our new Gemini photometry in the context of atmospheric models and brown dwarf populations. We conclude in $\S$\ref{sec:conclusion}.

\section{The Sample} 
\label{sec:sample}

\subsection{Target Selection for New Near-Infrared Photometry}
\label{sec:target_selection}

\citet{Meisner_2020a} identified $\sim$170 candidate cool brown dwarfs using  imaging data in filters centered near $3.4~\mu$m and $4.6~\mu$m: WISE's W1 and W2 filters \citep{Cutri_2012,Wright_2010}, and $Spitzer$ IRAC [3.6] = ch1 and [4.5] = ch2 \citep{Fazio_2004,Werner_2004}. NIR photometry was obtained for many of the targets either via new observations or from the 2MASS, UKIDSS, and VISTA sky survey catalogs \citep{Skrutskie_2006, Irwin_04, Hewett_2006, Lawrence_2007, Casali_2007, Hambly_2008, Dye_2018, Cross_2012, McMahon_2013}. \citet{Meisner_2020a}
noted a distinct subpopulation with anomalously large $J-$[4.5] colors for their relatively blue $Spitzer$ colors. Five objects were identified as members of this population:  CWISEP J015613.24+325526.6 (CWISEP 0156$+$3255 for short), CWISEP J050521.29$-$591311.7 (CWISEP 0505$-$5913 for short), CWISEP J070055.19+783834.0 (CWISEP 0700$+$7838 for short), CWISEP J090536.35+740009.1 (CWISEP 0905$+$7400 for short), and WISEA J153429.75$-$104303.3 (WISEA 1534$-$1043 for short). WISEA 1534$-$1043 was highlighted as especially unusual in a later paper by \citet{Kirkpatrick_2021b}, which explored a variety of possible scenarios for explaining this object, concluding that it is likely metal poor, and perhaps the first Y-type subdwarf. This work provides new near-infrared photometry for the three sources with the reddest $J - $[4.5] color among these five candidates, suggesting they may be the colder sources: CWISEP 0156$+$3255, CWISEP 0505$-$5913, and WISEA 1534$-$1043. Of the remaining sources, \citet{Meisner_2020a} provide a $J$-band detection for CWISEP 0905$+$7400, and a lower limit for the $J$-band magnitude of CWISEP 0700$+$7838.

\citet{Meisner_2020b} provides follow-up \textit{Spitzer} mid-infrared imaging of 95 candidate cool brown dwarfs. In that work, the authors highlight 
WISEA 1553$+$6933 as a T-type subdwarf candidate with an exceedingly high tangential velocity estimate. In more recent work, \citet{Meisner_2021} and \citet{Brooks_2022} present two and one additional candidate ``extreme T-type subdwarfs'' (esdTs; ultracool dwarfs with $T_{\rm{eff}} \lesssim 1400$~K and [m/H] $\le -1$ dex), respectively. We target all four of these sources here for NIR imaging: CWISE J052306.42$-$015355.4 (CWISE 0523$-$0153 for short), CWISE J073844.52$-$664334.6 (CWISE 0738$-$6643 for short), WISEA J155349.96$+$693355.2 (WISEA 1553+6933 for short), and CWISE J221706.28$-$145437.6 (CWISE 2217$-$1454 for short). 

We also include in the observational sample a moderately metal-poor brown dwarf CWISE J021948.68$+$351845.3 (CWISE 0219+3518 for short), which is a companion to the M dwarf Ross 19, and  can therefore be used as a benchmark. \citet{Schneider_2021}
discovered the companion, and find that the system has [Fe/H] $= -0.40 \pm 0.12$ with an age of around 7 Gyr.

In summary, new near-infrared photometry was obtained for eight targets:  CWISEP 0156$+$3255, CWISE 0219$+$3518, CWISEP 0505$-$5913, 
CWISE 0523$-$0153, CWISE 0738$-$6643, WISEA 1534$-$1043, WISEA 1553$+$6933, and CWISE 2217$-$1454.

\subsection{Other Sources}
\label{sec:other_sources}

An additional benchmark is used in this work, taking photometry from the literature: the T8 subdwarf WISE 2005$+$5424, also known as  Wolf 1130C.  This brown dwarf is a companion to a binary system composed of a subdwarf M star and a white dwarf, with an age around 10 Gyr \citep{Mace_2018}, and [m/H] $= -0.75$ \citep{Kesseli_2019}. 

Two other sources are highlighted here, using photometry taken from the literature:  WISEA J041451.67$-$585456.7 (WISEA 0414$-$5854 for short) and WISEA J181006.18$-$101000.5 (WISEA 1810$-$1010 for short). These two objects were identified by \citet{Schneider_2020} as possibly the first known esdTs. WISEA 1810$-$1010 has been found by \cite{Lodieu_2022} to lie at a surprisingly nearby distance of just 8.9$^{+0.7}_{-0.6}$ pc and potentially have a very low metallicity [Fe/H] $\approx$ $-$1.5 dex. We synthesized $Y$- and $H$-band photometry for WISEA 0414$-$5854 based on its spectrum from \cite{Schneider_2020}, anchored to its measured $J$-band magnitude. We find $Y_{MKO} = 20.32 \pm 0.20$ mag and $H_{MKO} = 19.45 \pm 0.15$ mag for WISEA 0414$-$5854. We did not synthesize $K$-band photometry for WISEA 0414$-$5854, as its available $K$-band spectrum is extremely faint and noisy. We similarly synthesized $Y$-band photometry for WISEA 1810$-$1010 based on its spectrum from \cite{Schneider_2020} and \cite{Lodieu_2022}, finding $Y_{MKO} = 18.01 \pm 0.08$ mag. The uncertainty in the synthesized magnitudes is estimated from the noise in the spectrum, combined in quadrature with the uncertainty in the anchor photometry.

We further highlight the thick disk early T dwarf candidate WISE J210529.08$-$623558.7 \citep[hereafter WISE 2105$-$6235;][]{Luhman_PlanetX} by computing synthetic $Y$-band and $H$-band photometry based on its spectrum published in \cite{Luhman_Sheppard_2014}. We obtain $Y_{MKO}$ = 17.96 $\pm$ 0.03 mag and $H_{MKO}$ = 15.91 $\pm$ 0.03 mag for WISE 2105$-$6235.

\section{New Near-Infrared Photometry}
\label{sec:observations}

\subsection{Filter Selection}

All filters used in this work are defined by the Maunakea Observatories filter specifications \citep[MKO;][]{Tokunaga_2002}. $J$-band imaging was obtained for six of the eight targets. For the four brighter sources within this group, we also obtained $H$-band imaging, and for two of those we obtained $K$-band imaging. The two remaining targets, for which $J$ had previously been measured,  were selected for $Y$-band imaging to explore the potential of the $Y$-band as a metallicity indicator \citep[e.g.,][]{Leggett_2017, Zhang_2019}.

Table \ref{tab:phot} lists the eight targets and the filters used in each observation.

\subsection{Gemini Observations}

Both Gemini North and Gemini South Observatories were used for this work. At Gemini North, GNIRS \citep{Elias_2006} was used for imaging, via program GN-2022A-Q-326. At Gemini South, Flamingos-2 \citep{Eikenberry_2006} was used for imaging, via program GS-2022A-Q-246. All data were taken in photometric conditions with FWHM around 1$''$ at both sites, except for the faintest target WISEA 1534$-$1043, which required better seeing of $0\farcs 6$.

When employing GNIRS for photometry, the small ``key-hole" field has to be used, which measures approximately 20$''$ by 30$''$ on sky. A 9-step dither pattern with 3$''$ offsets was used. For all GNIRS data, the brown dwarf photometry was determined using an aperture of diameter $1\farcs 5$, and the aperture corrections (to match the larger aperture used for the calibrators) were determined from brighter 15th to 17th magnitude stars in the field of each brown dwarf.  For the $Y$-band imaging, all data were taken on the same night. The images were flux-calibrated by observing the UKIRT Faint Standard FS111
\footnote{\url{https://www.gemini.edu/observing/resources/near-ir-resources/photometry/ukirt-standards}} on this same night. The standard and the two brown dwarfs were all observed at an airmass around 1.2.  The $J$-band image of CWISE 0523$-$0153 was flux calibrated using a 15th magnitude VISTA Hemisphere Survey \citep[VHS;][]{McMahon_2013} star in the field of the brown dwarf; the same aperture was used to measure the photometry for both the brown dwarf and survey star. The images for WISEA 1553$+$6933 were flux calibrated by observing the UKIRT Faint Standard FS139\footnote{\url{https://www.gemini.edu/observing/resources/near-ir-resources/photometry/ukirt-standards}} on the same nights as the brown dwarf. The standard star was observed at an airmass of around 1.3, and the brown dwarf  at an airmass of around 1.6. No corrections were made for the small airmass difference, as the atmospheric extinction though these MKO filters is small\footnote{\url{https://www.gemini.edu/observing/telescopes-and-sites/sites}}. 

Flamingos-2 at Gemini South offers a large $6'$ field of view for imaging. A 9-step dither pattern with 5$''$ offsets was used. 
The observations of CWISEP 0505$-$5913 and CWISE 0738$-$6643 were flux calibrated using 15 to 30 VHS stars of 14th to 18th magnitude in their respective fields. The VHS photometry included $J$ and $Ks$ values only; $H$ was estimated from $J - Ks$ (which ranged from 0.3 to 0.7) using the stellar colors given by \citet{Covey_2007}. For CWISEP 0505$-$5913 the standard deviations in the $J$ and $H$ zeropoints were 0.03 and 0.04 magnitudes, respectively. For CWISE 0738$-$6643  the standard deviations in the $J$, $H$ and $Ks$ zeropoints were 0.05, 0.04 and 0.03 magnitudes, respectively.  All photometry was determined using an aperture of diameter $1\farcs 8$. The CWISEP 0505$-$5913 field was observed at an airmass of 1.3 and the CWISE 0738$-$6643 field at an airmass of 1.6. For CWISE 2217$-$1454, no survey data overlapped with the brown dwarf field, and observations were made of the nearby UKIDSS Deep eXtra-Galactic Survey \citep{Swinbank_2013} field around the very faint standard VFS72, identified in \citet{Leggett_2020}. Thirteen 15th to 17th magnitude stars in this field were used to calibrate the CWISE 2217$-$1454 $J$- and $H$-band imaging data. The standard deviations in both $J$ and $H$ zeropoints were 0.03  magnitudes. Aperture corrections were determined from stars in the CWISE 2217$-$1454 field.

\begin{figure*}
\begin{center}
\plotone{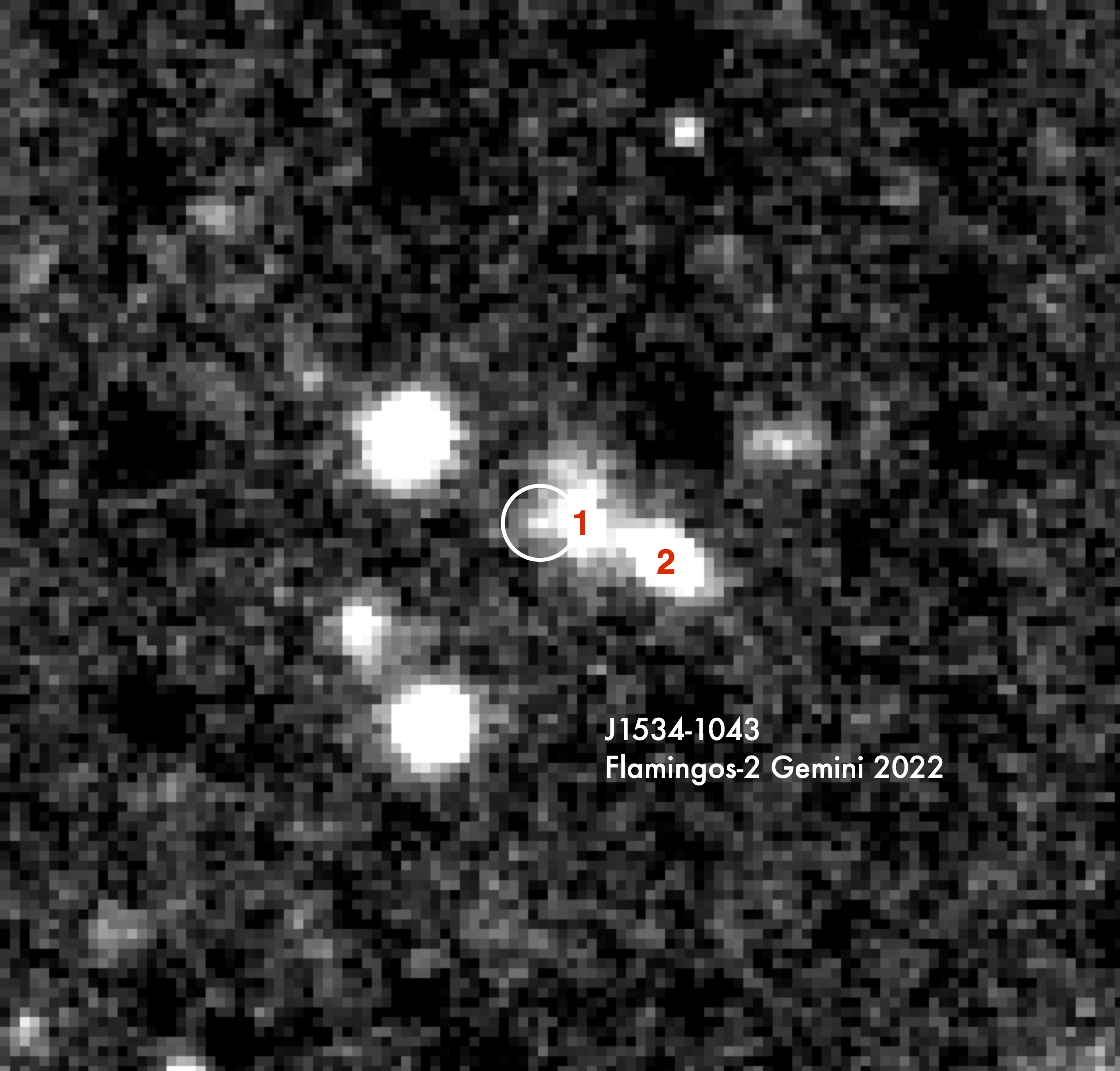}
\caption{Flamingos-2 stacked $J$-band image of WISEA 1534$-$1043. North is up and East is to the left. The image is approximately 20$''$ on a side. The white circle centered on the location of WISEA 1534$-$1043 is 1$\farcs$4 in diameter. Two relatively bright and nearby galaxies are labeled as ``1'' and ``2'' in red text.
\label{fig:1534_detection}}
\end{center}
\end{figure*}

The faintest target in our sample is the unusual WISEA 1534$-$1043. \citet{Kirkpatrick_2021b} determined a lower limit of $J_{MKO} > 23.8$ mag using the MOSFIRE instrument \citep{McLean_2012} at the W. M. Keck Observatory, and detected the source using the Hubble Space Telescope with F110W $= 24.70 \pm 0.08$ mag. We used 10 hours of queue time at Gemini South, in photometric conditions with seeing of $0\farcs 6$, to detect this target at $J_{MKO} = 24.5 \pm 0.3$ mag. The target was observed on two nights, at an airmass between 1.1 and 1.7.  The imaging was flux calibrated using 11 VISTA VHS  stars ranging in $J$ magnitude from 15.7 to 17.2. The standard deviation in the $J$  zeropoint was 0.016  magnitudes. The photometry was determined using an aperture of diameter $0\farcs 7$, and aperture corrections were determined from stars in the  field. WISEA 1534$-$1043 has a large proper motion of  $2\farcs 69$ yr$^{-1}$ in a south-westerly direction \citep{Kirkpatrick_2021b}, moving it very close to a group of background sources as shown in Figure \ref{fig:1534_detection}; the brown dwarf will once again be in a clean sky region as of approximately 2023 April (this statement is dependent in detail on the angular resolution and choice of bandpass). The predicted Gemini/Flamingos-2 position of WISEA 1534$-$1043 agrees at the better than $1$ pixel ($0.18''$) level with the location of the central source (white circle) in Figure \ref{fig:1534_detection}, hence there is no ambiguity about our identification of the correct counterpart. The quoted 0.3 mag uncertainty is predominantly due to the uncertainty of the sky background due to the crowded field. As a cross-check on the accuracy of our measured WISEA 1534$-$1043 $J$-band flux, we modeled each of the two galaxies labeled ``1'' and ``2'' in Figure \ref{fig:1534_detection} as a two-dimensional Gaussian light distribution, then subtracted these models to obtain an isolated view of WISEA 1534$-$1043. Performing aperture photometry on this galaxy-subtracted image, again using an aperture of diameter $0\farcs 7$, we find a flux measurement consistent with that in the unsubtracted image at the 3\% level. This is well within our quoted $\pm 0.3$ mag uncertainty on the WISEA 1534$-$1043 $J$-band magnitude.

\begin{deluxetable*}{lcccccc}
\tabletypesize{\small}
\tablecaption{Gemini Imaging Observations}
\tablehead{
\colhead{Target} & \colhead{Discovery} & \colhead{Filter} &
 \colhead{Magnitude} & \colhead{Instrument} & \colhead{Total On-} & \colhead{Date UT}  \\
\colhead{WISE RA$\pm$Dec} & \colhead{Reference\tablenotemark{a}} & \colhead{(MKO)}
& \colhead{(Vega)} & \colhead{} & \colhead{Source Time (hr)} & \colhead{YYYYMMDD}   }
\startdata 
015613.24$+$325526.6 & 1 &  Y & 21.94 $\pm$ 0.06 & GNIRS &  0.60   &  20220206 \\
021948.68$+$351845.3\tablenotemark{b} & 2 & Y & 21.86 $\pm$ 0.06 & GNIRS &  0.60   &  20220206 \\
050521.29$-$591311.7 & 1 & J & 20.74 $\pm$ 0.07 & Flamingos-2 &  0.10  & 20220201  \\
050521.29$-$591311.7 & 1 & H & 20.62 $\pm$ 0.08 & Flamingos-2 & 0.14  & 20220203  \\
052306.42$-$015355.4 & 3 & J & 19.14 $\pm$ 0.05 & GNIRS & 0.08    &  20220112 \\
073844.52$-$664334.6 & 4 & J & 21.37 $\pm$ 0.14 & Flamingos-2 & 0.21 & 20220128 \\
073844.52$-$664334.6 & 4 & H & 20.73 $\pm$ 0.14 & Flamingos-2 & 0.08 & 20220128 \\
073844.52$-$664334.6 & 4 & Ks\tablenotemark{c} & 21.44 $\pm$ 0.34 & Flamingos-2 & 0.11 & 20220302 \\
153429.75$-$104303.3 & 1 & J & 24.5 $\pm$ 0.3 & Flamingos-2 & 6.43 & 20220420, 20220611 \\ 
155349.96$+$693355.2 & 5 & J & 19.17 $\pm$ 0.03 & GNIRS & 0.04 &  20220221 \\
155349.96$+$693355.2 & 5 & H & 18.87 $\pm$ 0.05 & GNIRS &  0.04 &  20220221 \\
155349.96$+$693355.2 & 5 & K & 19.24 $\pm$ 0.03 & GNIRS & 0.53   &  20220214 \\
221706.28$-$145437.6 & 4 & J & 20.66 $\pm$ 0.02 & Flamingos-2 &  0.04 & 20220514 \\
221706.28$-$145437.6 & 4 & H & 20.66 $\pm$ 0.06 & Flamingos-2 &  0.10 & 20220514 \\
\enddata
\tablenotetext{a}{References - 1: \citet{Meisner_2020a}, 2: \citet{Schneider_2021}, 3: \citet{Brooks_2022}; 4: \citet{Meisner_2021}, 5: \citet{Meisner_2020b}.}
\tablenotetext{b}{Also known as Ross 19B \citep{Schneider_2021}.}
\tablenotetext{c}{\citet{Stephens_2004} show that $K_s - K_{MKO} = -0.10 \pm 0.05$ for mid- to late-T types, and we adopt $K_{MKO} = 21.54 \pm 0.34$ for CWISE 0738$-$6643.}
\label{tab:phot}
\end{deluxetable*}

\subsection{Gemini  Data Reduction}

The DRAGONS software package \citep{Labrie_2019} was used to reduce all of our new imaging data obtained at Gemini Observatory for this work. DRAGONS documentation is available at: \url{https://dragons.readthedocs.io/en/stable/}.

For Gemini's infrared cameras, DRAGONS performs these initial steps: a non-linearity correction is applied; counts are converted from data numbers to electrons; bad pixel masks are applied; and the read and Poisson noise are added to the FITS extension which carries the variance information. Multiple dark observations are stacked to create a master dark. A master flat is created from multiple lamps-on and lamps-off observations; the flat is normalized and thresholded for out-of-range values. 

Science data is divided by the appropriate flat field for filter and read mode. The sky contribution is determined for each pointing using the images taken at other positions in the dither pattern. The sky is then subtracted from each science image. Point sources are detected in each image, and these are used to align and stack the data set for each object. Each sky-subtracted image in the stack is numerically scaled based on the background signal, by factors typically $< 5$\%, to produce a final image.  We used simple aperture photometry to measure magnitudes from the processed images, as described in the previous section.

\section{Discussion}
\label{sec:discussion}

\subsection{Old/Cold Sample in the Context of the Broader Brown Dwarf Population}
\label{sec:pop_diagrams}

\begin{figure*}
\begin{center}
\vskip -0.5in
\plotone{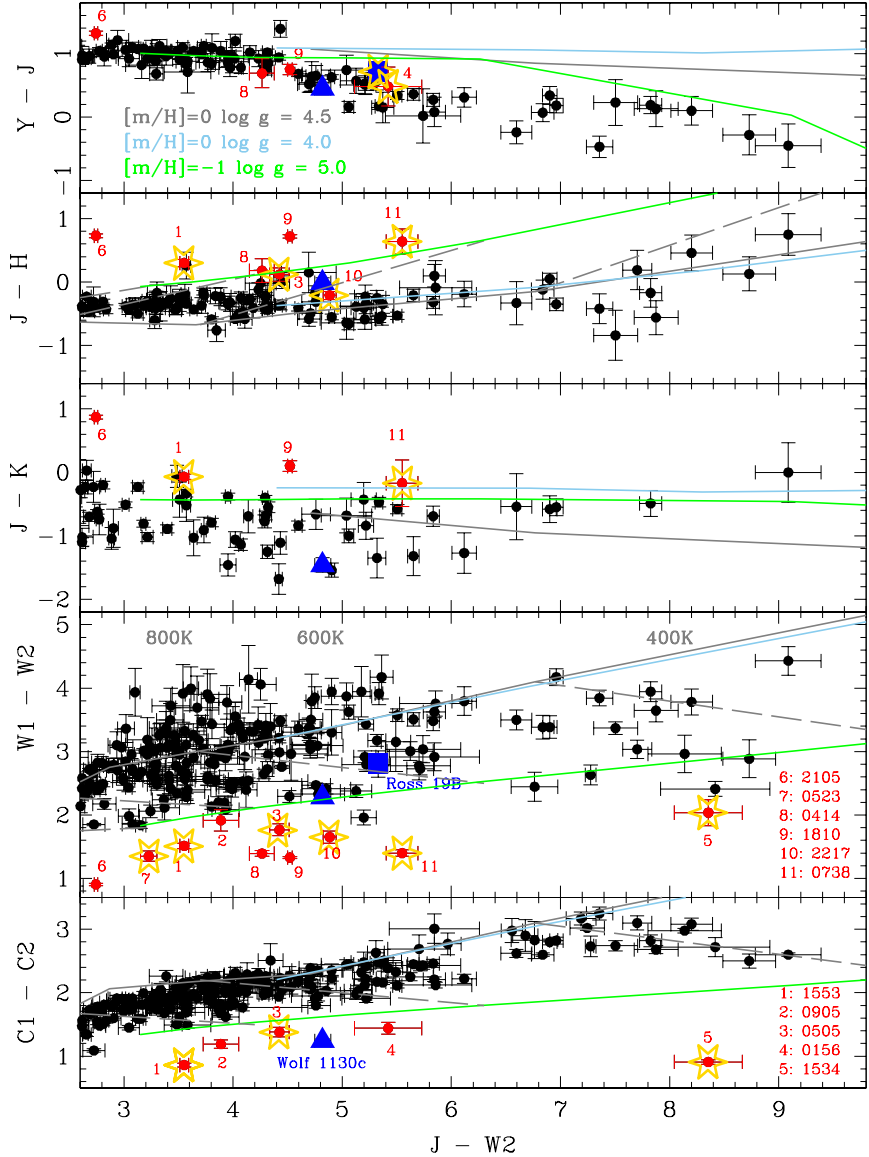}
\vskip -0.2in
\caption{Color-color plots for T and Y dwarfs. The upwards blue triangle (blue square) is the metal-poor benchmark T dwarf Wolf 1130C (Ross 19B). These two benchmark systems have metallicities of $-0.75$ and $-0.40$ dex, respectively (see $\S$\ref{sec:target_selection} and $\S$\ref{sec:other_sources}). Red points are (candidate) metal-poor brown dwarfs, identified by red text annotations as specified in the bottom two panels. Red points with surrounding yellow stars represent objects for which we present new Gemini photometry. The solid gray and solid light blue lines are model sequences with [m/H] = 0. The solid green line is a model sequence with [M/H] = $-1$. Note that all three model tracks have different log($g$) values, as indicated in the plot annotations. These model tracks have been generated from the tuned-adiabat atmospheric models presented by \cite{Leggett_2021}. Dashed gray lines in the $J$$-$$H$, W1$-$W2, and ch1$-$ch2 panels are lines of approximately constant $T_{\rm eff}$, from left to right: 800, 600, and 400 K.
\label{fig:color_color_sandy}}
\end{center}
\end{figure*}

With this new Gemini photometry in hand, we can place our sample of eight old/cold targets in context with the broader population of cool brown dwarfs. Metal-poor ultracool dwarfs have been noted as color outliers by many prior studies \citep[e.g.,][]{Logsdon_2018, Mace_2013a, Mace_2013b}, and we therefore sought to visualize our sample via a series of color-color and color-magnitude diagrams. Figure \ref{fig:color_color_sandy} shows color-color plots for each of $Y$$-$$J$, $J$$-$$H$, $J$$-$$K$, W1$-$W2, and \textit{Spitzer} ch1$-$ch2 as a function of $J$$-$W2. Also overplotted within each panel are [m/H] = 0, log($g$) = 4, [m/H] = 0, log($g$) = 4.5, and [m/H] = $-1$, log($g$) = 5 model tracks (\citealt{Tremblin_2015,Phillips_2020,Leggett_2021}; solid light blue, gray, and green lines, respectively). In the top ($Y$$-$$J$) panel of Figure \ref{fig:color_color_sandy}, the confirmed and candidate metal-poor objects (all having $J$$-$W2 $<$ 6 mag) do not stand out strongly from the ``normal'' brown dwarf sequence. In this same $J$$-$W2 $<$ 6 mag color range, the [m/H] = $-$1, 0 model tracks are likewise not majorly differentiated from one another, though they do diverge in $Y$$-$$J$ color toward redder $J$$-$W2. The $J$$-$$H$ panel of Figure \ref{fig:color_color_sandy} suggests that metal-poor brown dwarfs have a moderate $J$$-$$H$ excess with respect to ``normal'' brown dwarfs, in reasonable agreement with the model tracks. This may indicate that future photometric follow-up of metal-poor brown dwarf candidates could benefit from prioritizing $H$-band over $Y$-band, though $Y$-band may still be highly valuable for the very coldest metal-poor candidates like WISEA 1534$-$1043 (see $\S$\ref{sec:w1534}).

No clear-cut narrative emerges as of yet from the $J$$-$$K$ panel of Figure \ref{fig:color_color_sandy}. Four (candidate) metal-poor objects (CWISE 0738$-$6643, WISEA 1553+6933, WISEA 1810$-$1010, and WISE 2105$-$6235 (from \citealt{Luhman_PlanetX}) have $J$$-$$K$ color $\sim$0.5-1.5 mag redder than typical brown dwarfs of similar $J$$-$W2 colors, whereas the benchmark Wolf 1130C (blue triangle) is on the blue side of typical in $J$$-$$K$ color. In order to draw firm conclusions, more $K$-band spectroscopy or photometry must be collected for metal-poor brown dwarfs, particularly given that H$_{2}$ opacity prominent at $K$-band is sensitive to both metallicity and gravity; if there is a range in gravity (i.e., mass/age) at a given $T_{\rm{eff}}$ then that would obfuscate the effect of metallicity. 

The bottom two panels of Figure \ref{fig:color_color_sandy} show W1$-$W2 and ch1$-$ch2 color as a function of $J$$-$W2. The (candidate) metal-poor population is  well separated from the ``normal'' brown dwarf sequence in both of these panels, with W1$-$W2 and ch1$-$ch2 colors both significantly bluer for metal-poor objects at fixed $J$$-$W2 color. The models are also in good agreement with the data regarding this trend. This finding reinforces the suggestion of \cite{Meisner_2021} that metal-poor T dwarf candidates can be selected by identifying objects that have relatively blue W1$-$W2 for their $J$$-$W2 color. The separation between metal-poor candidates and the broader population is somewhat cleaner in terms of ch1$-$ch2 than W1$-$W2, which may arise from the superior sensitivity/resolution of \textit{Spitzer} as compared to WISE, or differences between how the W1 and ch1 bandpasses integrate against the reduction of 3-4~$\mu$m methane absorption toward lower metallicities. The Appendix compares the WISE and \textit{Spitzer} photometry for our sample, and also illustrates the importance of using the proper motion corrected magnitudes given in CatWISE2020 \citep{Marocco_2021}.

\begin{figure*}
\begin{center}
\vskip -0.8in
\plotone{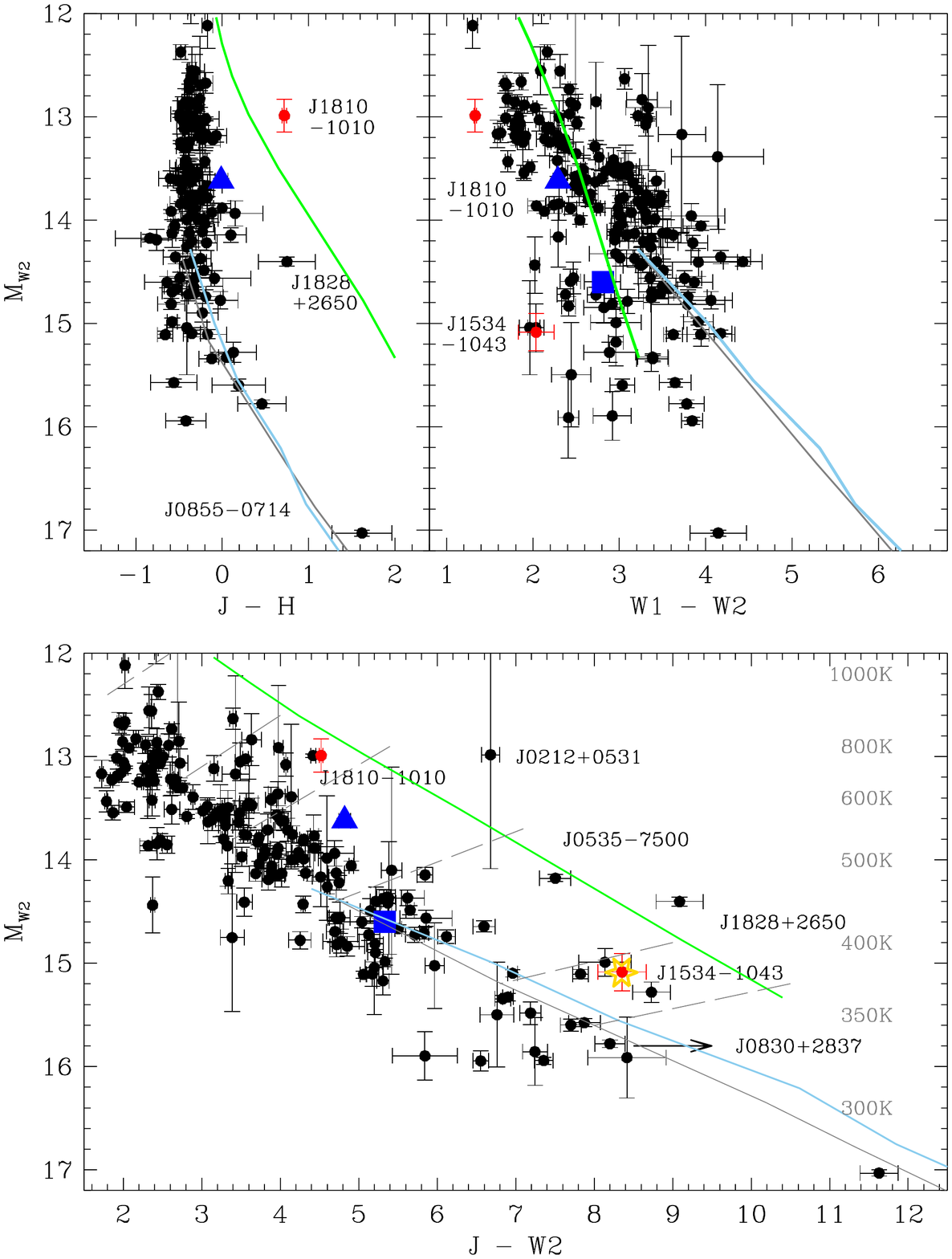}
\vskip -0.2in
\caption{Color-magnitude plots for T and Y dwarfs. The upwards blue triangle (blue square) is the metal-poor benchmark T dwarf Wolf 1130C (Ross 19B). 
Outliers are identified. We present new $J$-band photometry for WISEA 1534$-$1043 in this work (red dot with surrounding yellow star). Red dots without a yellow star are data points from the prior literature for metal-poor brown dwarf candidates. The black arrow in the bottom panel illustrates the lower limit on $J - $W2 from \citet{Bardalez_2020} for WISEA 0830+2837. J0212+0531 is CWISEP J021243.55+053147.2 \citep{Meisner_2020a,Leggett_2021}. J0535$-$7500 is WISE J053516.80$-$750024.9 \citep{Kirkpatrick_2012}.  J1828+2650 is WISEPA J182831.08+265037.8. \citep{Cushing_2011,Kirkpatrick_2011}. \label{fig:color_mag_sandy}}
\end{center}
\end{figure*}

Figure \ref{fig:color_mag_sandy} shows W2 absolute magnitude ($M_{W2}$) as a function of $J$$-$$H$, W1$-$W2, and $J$$-$W2 for our Gemini sample along with a broader set of literature T and Y dwarfs. Note that many members of our old/cold sample lack trigonometric parallaxes and hence cannot be included in Figure \ref{fig:color_mag_sandy}. Overplotted within each panel are the same three model tracks from Figure \ref{fig:color_color_sandy}. The models predict that $M_{W2}$ is brighter for [m/H] = $-1$ than for [m/H] = 0 at fixed $J$$-$$H$ color. WISEA 1810$-$1010 aligns with this trend assuming it has a metallicity somewhat below $-1$. Wolf 1130C falls in between the [m/H] = $-1$, 0 tracks in the $J$$-$$H$ panel, consistent with its benchmark metallicity of $-0.75$ dex. The [M/H] = $-1$, 0 model tracks in $M_{W2}$ versus W1$-$W2 have little overlap along the WISE color axis, complicating interpretation. Nevertheless, it is clear that metal-poor candidates WISEA 1534$-$1043 and WISEA 1810$-$1010 are outliers compared to the general brown dwarf population, being relatively blue in W1$-$W2 at fixed $M_{W2}$, or alternatively fainter in $M_{W2}$ at fixed W1$-$W2. Wolf 1130C and Ross 19B are more consistent with the broader population in the $M_{W2}$ versus W1$-$W2 panel, but both deviate at least slightly in the same sense as WISEA 1534$-$1043 and WISEA 1810$-$1010. The bottom panel of Figure \ref{fig:color_mag_sandy} shows $M_{W2}$ as a function of $J$$-$W2 color. At fixed $J$$-$W2, the [m/H] = 0 track corresponds to a cooler temperature and hence fainter $M_{W2}$ than the [m/H] = $-1$ track. The (candidate) metal-poor objects WISEA 1534$-$1043, WISEA 1810$-$1010, and Wolf 1130C all fall in between the [m/H] = $-1$, 0 model tracks in the bottom panel of Figure \ref{fig:color_mag_sandy}, suggesting metallicities in the range $-1 < $ [m/H] $< 0$. However, binarity rather than metallicity could be an alternative effect pushing these objects toward brighter absolute magnitudes at fixed color. Ross 19B appears consistent with [M/H] $\approx$ 0 in terms of $M_{W2}$ verus $J$$-$W2.

\begin{figure*}
\begin{center}
\includegraphics[width=6.0in]{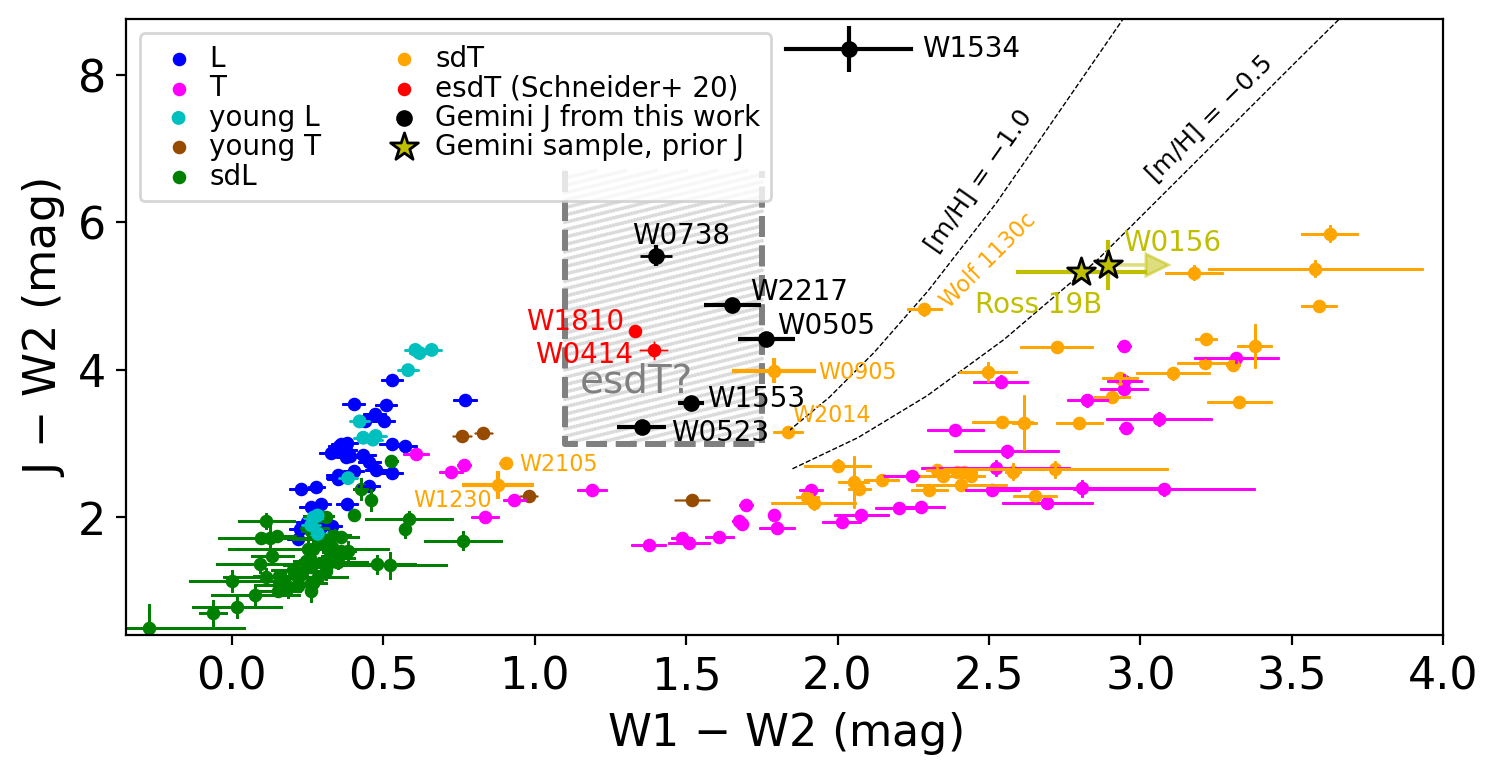}
\caption{Color-color diagram (adapted from \citealt{Meisner_2021}) highlighting candidate extreme T type subdwarfs and updated to reflect the $J$-band results of our Gemini follow-up imaging campaign (black dots). The other samples shown are L dwarfs (blue), young L dwarfs (cyan), L subdwarfs (green), ``normal'' T dwarfs (magenta), young T dwarfs (brown), T subdwarfs (orange), esdT candidates from \citealt{Schneider_2020} (red; W0414 = WISEA 0414$–$5854, W1810 = WISEA 1810$–$1010), and members of our present Gemini sample with $J$ magnitudes from the prior literature (yellow stars). Note the location of WISEA 1534$-$1043 near the very upper middle. The shaded gray box indicates the fiducial esdT candidate color-color selection proposed in \cite{Meisner_2021}. Dotted black lines show our model tracks for [m/H] = 0 and [m/H] = $-0.5$ (labeled in black text near the upper right end of each track). Both of these model tracks have log($g$) = 5, log($K_{zz}$) = 7, and $\gamma = 1.25$ \citep{Tremblin_2015}. CWISE 0738$-$6643 (labeled W0738) stands out as unusually red in $J$$-$W2 for its modest W1$-$W2 color, even among the sample of  esdT candidates. CWISEP J050521.29$-$591311.7 (labeled W0505), CWISEP 0905+7400 \citep[labeled W0905;][]{Meisner_2020a}, and 
WISE J201404.13+042408.5 \citep[labeled W2014;][]{Kirkpatrick_2012} are also individually labeled because they fall just outside of the fiducial esdT color-color selection box. The early T subdwarf candidate labeled W1230 is CWISE J123041.80+380140.9 from \cite{Kota_2022}.
\label{fig:color_color_aaron}}
\end{center}
\end{figure*}

Figure \ref{fig:color_color_aaron} shows a $J$$-$W2 versus W1$-$W2 color-color diagram for our Gemini sample in an even broader context of various LTY brown dwarf populations. [m/H] = $-0.5$ and [m/H] = $-1$ model tracks are overplotted as dotted black lines. Ross 19B and CWISEP 0156+3255 land at similar locations within this color-color diagram, very close to the [m/H] = $-0.5$ track\footnote{Although CWISEP 0156+3255 is shown in Figure \ref{fig:color_color_aaron} with a W1$-$W2 color limit based on a claimed W1 non-detection in CatWISE2020, the CatWISE Preliminary Catalog listed it as detected in both W1 and W2, with a color of 2.749 $\pm$ 0.424 mag, suggesting that the placement of our arrow is near to the actual color-color location at which this object lands.}. This aligns well with the fact that we believe the Ross 19 system to have [m/H] $\approx$ $-0.4$ based on detailed characterization of its M dwarf primary \citep{Schneider_2021}. Ross 19B and CWISEP 0156+3255 also fall in a similar location within the $Y$$-$$J$ versus $J$$-$W2 color-color panel of Figure \ref{fig:color_color_sandy}. These considerations suggest that CWISEP 0156+3255 may have a metallicity in the range of $-0.4$ to $-0.5$ dex, reinforcing its status as a T-type subdwarf candidate. It would be valuable to obtain a higher S/N $J$-band detection of CWISEP 0156+3255, as the \cite{Meisner_2020a} $J$-band measurement has a large 0.3 mag uncertainty indicating a marginal detection.

In the bottom two panels of Figure \ref{fig:color_color_sandy} and in Figure \ref{fig:color_color_aaron}, WISEA 1534$-$1043 still stands far apart from the remainder of our old/cold sample. We can see that CWISE 0738$-$6643 represents the object closest to `forming a bridge' between WISEA 1534$-$1043 and the esdT candidate population, in the sense that CWISE 0738$-$6643 has the largest $J$$-$W2 color among esdT candidates\footnote{We consider WISEA 1534$-$1043 to be more so an sdY candidate than an esdT candidate, as this object falls outside of the nominal esdT color-color selection box proposed in \cite{Meisner_2021} and shown in Figure \ref{fig:color_color_aaron}}.  The relatively large $J$$-$W2 color of CWISE 0738$-$6643 may be an indication that it is the coolest of the as-yet identified esdT candidates, making it a relatively high priority for additional future follow-up.

The second reddest esdT candidate in terms of $J$$-$W2 (i.e., potentially second coldest) is CWISE 2217$-$1454. Acquiring $K$-band photometry of CWISE 2217$-$1454 would allow us to search for the enhancement of H$_2$ collision induced absorption (CIA) previously seen in T-type subdwarfs \citep[e.g.,][]{Zhang_2019}, and gauge whether this effect is present in CWISE 2217$-$1454. CWISE 2217$-$1454 would be a relatively high leverage data point if added to the Figure \ref{fig:color_color_sandy} $J$$-$$K$ versus $J$$-$W2 diagram, as the model tracks predict that $J$$-$$K$ color diverges somewhat more with metallicity toward redder $J$$-$W2. Spectroscopic confirmation/metallicity are also needed for CWISE 0738$-$6643 and would aid in assessing whether blue or rather red $J$$-$$K$ color should be taken as an indicator of very low metallicity, $< -0.5$ dex, in T dwarfs.

Several objects land only slightly outside of the gray hatched esdT selection box drawn in Figure \ref{fig:color_color_aaron}: CWISEP 0505$-$5913, CWISEP 0905$+$7400, WISE J201404.13+042408.5 \citep{Kirkpatrick_2012}, and WISE 2105$-$6235. The former three fall just redward of the esdT selection box's right side W1$-$W2 = 1.75 mag boundary, while the latter is somewhat blueward of both the left boundary (W1$-$W$2 \ge 1.1$ mag) and lower boundary ($J$$-$W2 $\ge$ 3 mag). We caution that the nominal \cite{Meisner_2021} esdT color-color selection criteria were tentative/preliminary, such that these objects near the boundary also merit additional follow-up toward understanding the newly identified old/cold brown dwarf population(s). WISE 2105$-$6235 is unusual in that it falls blueward rather than redward of the nominal esdT selection in terms of W1$-$W2 color. We note that WISE 2105$-$6235 is contaminated by a background source during the pre-hibernation WISE mission, so that its W1 and W2 magnitudes from AllWISE and CatWISE are likely biased. Using the unTimely Catalog \citep{Meisner_2022}, which performs independent flux measurements per WISE sky pass, we find W1 = 14.82 $\pm$ 0.01 mag and W2 = 13.93 $\pm$ 0.02 mag, excluding WISE epochs earlier than year 2019.0. This curated unTimely photometry results in W1$-$W2 = $0.89 \pm 0.02$ mag and $J$$-$W2 = $2.71 \pm 0.02$ mag, which are quite similar to values obtained from CatWISE2020 and therefore would not shift WISE 2105$-$6235 into the esdT color-color region.

\begin{figure*}
\begin{center}
\includegraphics[width=6.0in]{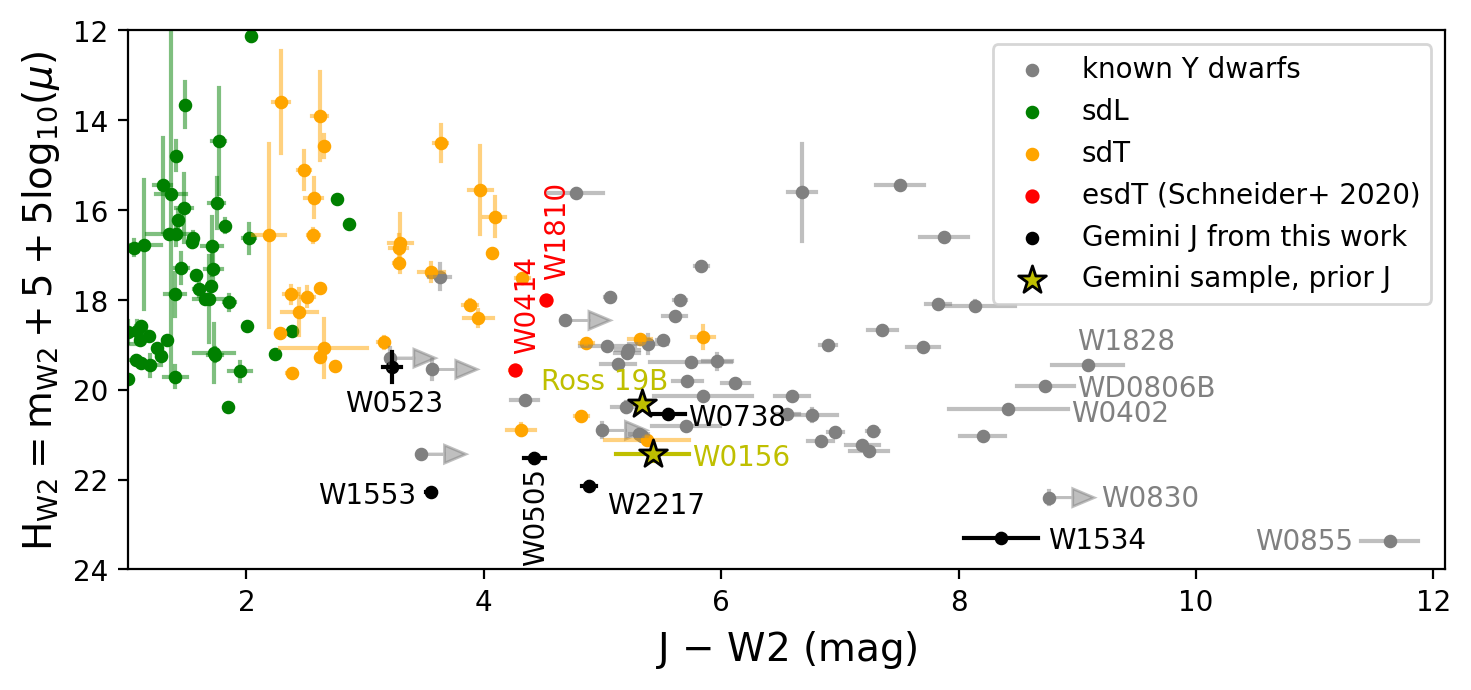}
\caption{Reduced proper motion diagram for members of our sample with Gemini $J$-band detections (black points), L subdwarfs (green), T subdwarfs (orange), esdT candidates from \citealt{Schneider_2020} (red; W0414 = WISEA 0414$–$5854, W1810 = WISEA 1810$–$1010), and known Y dwarfs (gray). Ross 19B and CWISEP 0156+3255 (labeled W0156) are indicated with yellow stars because they are members of our Gemini sample with $J$-band detections from the prior literature \citep{Meisner_2020a,Schneider_2021}. Some of the Y dwarfs which appear most extreme in this diagram are individually labeled: W0402 = CWISEP J040235.55$-$265145.4 \citep{Meisner_2020a}, WD0806B = WD 0806$-$661B \citep{Luhman_2011}, W0830 = WISEA 0830+2837 \citep{Bardalez_2020}, W0855 = WISE 0855$-$0714 \citep{Luhman_2014}, and W1828 = WISEPA J182831.08+265037.8. \citep{Cushing_2011,Kirkpatrick_2011}.
\label{fig:reduced_pm}}
\end{center}
\end{figure*}

Figure \ref{fig:reduced_pm} shows a reduced proper motion diagram \citep{Jones_1972} leveraging our new Gemini $J$-band photometry in combination with literature data for other L-type subdwarfs, T-type subdwarfs, candidate esdTs, and  Y dwarfs. Reduced proper motion (here given by $H_{W2} = m_{W2} + 5 + 5$log$_{10}\mu$, with $\mu$ in arcseconds per year) is a useful stand-in for absolute magnitude when a trigonometric parallax is unavailable, as is the case for most of the objects plotted in Figure \ref{fig:reduced_pm}. Reduced proper motion increases both with high kinematics (large $V_{tan}$) and with decreasing luminosity. Hence, Figure \ref{fig:reduced_pm} should tend to highlight the oldest/coldest brown dwarfs as having high $H_{W2}$ (i.e., landing toward the lower boundary of the reduced proper motion diagram) and high $J$$-$W2 \citep[which increases with decreasing temperature and decreasing metallicity e.g.,][]{Kirkpatrick_2011, Meisner_2021}. Among our Gemini sample, only Ross 19B and WISEA 1534$-$1043 have parallaxes available\footnote{In Figure \ref{fig:color_mag_sandy}, we have assigned the Ross 19 Gaia parallax to Ross 19B, on the premise that they are physically associated.}, so reduced proper motion is particularly valuable for providing another window into which old/cold sample members may be relatively extreme, in the sense of having unusually low luminosities and/or extremely high kinematics. WISEA 1534$-$1043 stands out in terms of high W2 reduced proper motion (exceeded only by WISE 0855$-$0714 among ultracool dwarfs) and red $J$$-$W2 color (exceeded only by a handful of the coolest known Y dwarfs; see $\S$\ref{sec:w1534} for further details). CWISE 0738$-$6643 and CWISE 2217$-$1454 also appear relatively extreme (relatively far toward the lower right) in this reduced proper motion diagram when compared to our remaining Gemini $J$-band targets (CWISEP
0505$-$5913, CWISE 0523$-$0153, and WISEA 1553+6933) suggesting once again that CWISE 0738$-$6643 and CWISE 2217$-$1454 may be the most extreme esdT candidates in terms of temperature, metallicity, and/or kinematics. However, in Figure \ref{fig:reduced_pm}, CWISE 0738$-$6643 lands very near Ross 19B while CWISE 2217$-$1454 falls in the same vicinity as CWISEP 0156+3255, two objects thought to be only moderately metal poor ([m/H] $\gtrsim$ $-0.5$). Though CWISE 0738$-$6643 and CWISE 2217$-$1454 may be superlative among candidate esdTs based on available photometric/astrometric indicators, near-infrared spectroscopy will be critical to determine whether their metallicities are truly extreme ([m/H] $\lesssim -1$) or more mildly metal poor ([m/H] $\gtrsim$ $-$0.5).

\subsection{The Temperature and Metallicity of WISEA 1534$-$1043}
\label{sec:w1534}

Our Gemini $J$-band detection of WISEA 1534$-$1043 is the first ever ground-based detection of this object, which had previously only been detected from space with WISE, \textit{Spitzer}, and Hubble Space Telescope (HST) in the F110W bandpass spanning the $\sim$0.9-1.4~$\mu$m wavelength range.

\begin{figure*}
\begin{center}
\includegraphics[width=5.5in]{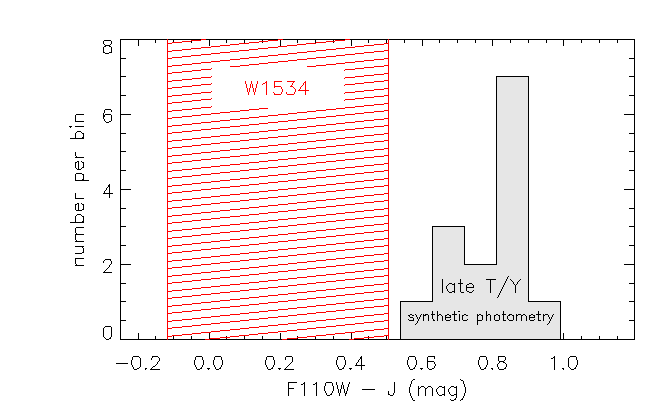}
\includegraphics[width=5.5in]{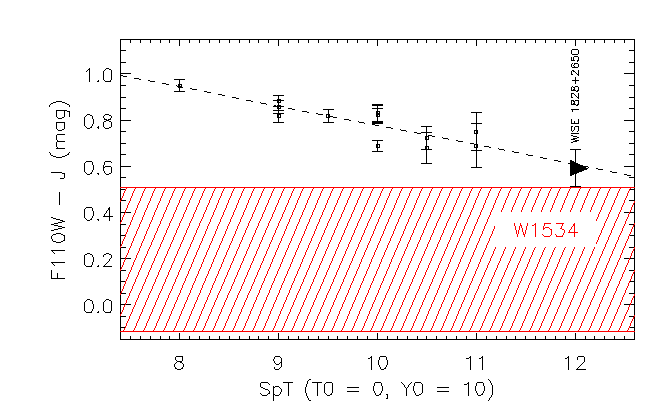}
\caption{Top: The F110W$-$$J$ color of WISEA 1534$-$1043 (red hatched regions) compared to F110W$-$$J$ color synthesized from Hubble Space Telescope spectroscopy of 14 late-T and Y dwarfs \citep[gray histogram;][]{Cushing_2011,Cushing_2014,Cushing_2021,Kirkpatrick_2012,Schneider_2015}. Bottom: F110W$-$$J$ color versus spectral type, for the same sample of 14 late-T and Y dwarfs. There is a roughly linear trend between F110W$-$$J$ color and numerical spectral type (T0 = 0, Y0 = 10). W1534 has a significantly bluer F110W$-$$J$ color than the other late T and Y dwarfs, such that the 1$\sigma$ upper envelope of its F110W$-$$J$ color remains bluer than the bluest F110W$-$$J$ color in the comparison sample. This indicates that WISEA 1534$-$1043 has an unusual spectrum in the 0.9-1.4~$\mu$m range relative to the known population of brown dwarfs currently thought to be similar in temperature. \label{fig:f110w_minus_j}}
\end{center}
\end{figure*}

Figure \ref{fig:f110w_minus_j} shows a histogram of F110W$-$$J$ color for late T and Y dwarfs for which it is possible to synthesize this color from archival Hubble Space Telescope near-infrared spectroscopy \citep{Cushing_2011,Cushing_2014,Cushing_2021,Kirkpatrick_2012,Schneider_2015}. As noted in \citet{Kirkpatrick_2021b}, the typical F110W$-$$J$ color for late T and Y dwarfs is $\sim$0.8 mag. Based on our new Gemini $J$-band detection, WISEA 1534$-$1043, on the other hand, has F110W$-$$J$ = 0.20 $\pm 0.31$ mag, indicated by the red hatched region in Figure \ref{fig:f110w_minus_j}. WISEA 1534$-$1043 is bluer than all other objects in our late-T/Y comparison sample with F110$-$$J$ color available. The next bluest object in F110W$-$$J$ is WISEPA J182831.08+265037.8 \citep[WISE 1828+2650 for short;][]{Cushing_2011,Kirkpatrick_2011}, itself an enigmatic Y-type dwarf presently thought to be perhaps a tight pair of two $\sim$325 K, $\sim$5~$M_J$ objects \citep{Leggett_2013,Cushing_2021b}. Still, WISEA 1534$-$1043 is so much bluer than WISE 1828+2650 in F110W$-J$ that the 1$\sigma$ color upper envelope for WISEA 1534$-$1043 remains lower than the central value of F110W$-J$ for WISE 1828+2650 (F110W$-J$ = $0.59 \pm 0.08$ mag). The Figure \ref{fig:f110w_minus_j} sample of comparison objects with synthetic F110W$-J$ available ranges from spectral type T8 to spectral type $\ge $Y2, the latter pertaining to WISE 1828+2650.

The bottom panel of Figure \ref{fig:f110w_minus_j} shows that there is a clear trend of decreasing F110W$-$$J$ color toward later spectral types in the T8-Y2 range. This trend could potentially hint at an extremely cold temperature for WISEA 1534$-$1043, but no such conclusion can be drawn without the ability to disentangle effects of temperature versus metallicity, both of which may tend to decrease F110W$-$$J$.

Based on the model tracks and associated iso-temperature (dashed gray) lines shown in Figures \ref{fig:color_color_sandy} and \ref{fig:color_mag_sandy}, WISEA 1534$-$1043 has photometry most consistent with $T_{\rm{eff}} \approx 400$-550~K and [m/H] $\lesssim$ $-0.5$ dex but perhaps significantly lower. The $J$$-$W2 versus W1$-$W2 panel of Figure \ref{fig:color_color_sandy} indicates that WISEA 1534$-$1043 has a metallicity $< -1$ dex, because it falls blueward of the [m/H] = $-1$ dex model track in terms of W1$-$W2 color. Linearly extrapolating the $J$$-$W2 versus W1$-$W2 iso-temperature contours suggests $T_{\rm{eff}} \approx 550$~K for WISEA 1534$-$1043. Our $M_{W2}$ versus W1$-$W2 color-magnitude plot (Figure \ref{fig:color_mag_sandy}) also suggests [m/H] $< -1$ dex for WISEA 1534$-$1043, as WISEA 1534$-$1043 falls below the [m/H] = $-1$ model track. A metallicity below $-1$ dex for WISEA 1534$-$1043 would be plausible given that this object displays kinematics characteristic of the Milky Way halo \citep{Kirkpatrick_2021b}.  On the other hand, our $M_{W2}$ versus $J$$-$W2 color-magnitude plot suggests a less extreme metallicity for WISEA 1534$-$1043, perhaps near the middle of the $-$1 $<$ [m/H] $<$ 0 range. Our $M_{W2}$ versus $J$$-$W2 color-magnitude diagram also indicates a temperature near 400~K or perhaps slightly lower for WISEA 1534$-$1043. Nevertheless, because we find temperatures as high as 550~K to be plausible for WISEA 1534$-$1043, while the T/Y transition occurs at $\approx 485$~K \citep{Leggett_2021}, we cannot yet determine whether WISEA 1534$-$1043 is the first Y-type subdwarf. Further complicating matters, the T/Y boundary could depend in detail on metallicity and/or undergo significant revision in light of new JWST spectroscopy extending to mid-IR wavelengths. JWST spectroscopy will be a critical and necessary step for unveiling the detailed physical properties of WISEA 1534$-$1043.

Our model-based $T_{\rm{eff}} = 400$-550 K estimate for WISEA 1534$-$1043 agrees well with a simplistic temperature estimate based on its \textit{Spitzer} ch2 absolute magnitude of $M_{ch2} = 14.707 \pm 0.17$ \citep{Kirkpatrick_2021b}. Using the $T_{\rm{eff}}(M_{ch2})$ polynomial relation of \cite{Kirkpatrick_2019} fit to the general $T_{\rm{eff}} \lesssim 1000$ K brown dwarf population, this ch2 absolute magnitude translates into an effective temperature estimate of $T_{\rm{eff}} = 453 \pm 77$ K, where the uncertainty is dominated by the 73 K scatter observed in the training sample relative to the polynomial model.

\begin{figure*}
\begin{center}
\includegraphics[width=6.0in]{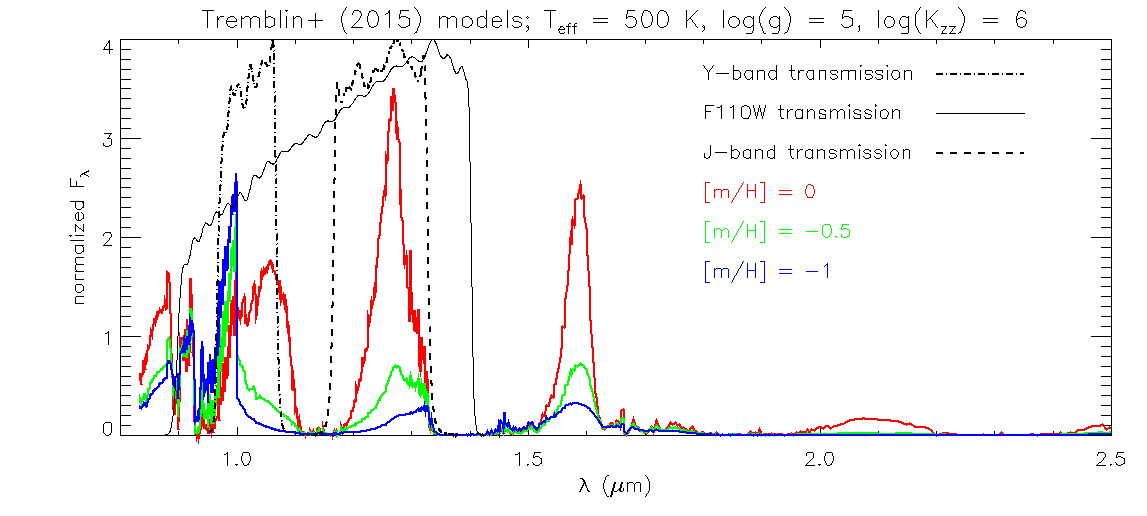}
\caption{A metallicity sequence of models from \cite{Tremblin_2015}. All models have $T_{\rm{eff}} = 500$ K, log($g$) = 5, and log($K_{zz}$) = 6. Red, green, and blue lines are the [m/H] = 0, $-0.5$, and $-1$ dex models, respectively. All models are normalized to unity between 0.91 and 0.925 microns. The dot-dashed black line shows the Gemini/Flamingos-2 $Y$-band transmission, the thin solid black line shows the F110W transmission, and the dashed black line provides the MKO $J$-band transmission. These transmission curves are displayed here with normalizations such that each has a maximum value of 4. As metallicity decreases, the $J$-band peak is dramatically suppressed such that a bluer color arises within the 0.9-1.4~$\mu$m wavelength range. Thus, the \cite{Tremblin_2015} models appear qualitatively in alignment with the relatively blue F110W$-$$J$ color of WISEA 1534$-$1043.\label{fig:model_spectra}}
\end{center}
\end{figure*}

Figure \ref{fig:model_spectra} shows the evolution of \cite{Tremblin_2015} model spectra for $T_{\rm{eff}}$ = 500 K as metallicity ranges between [m/H] = $-1$ dex (blue line) and [m/H] = 0 (red line). In the 0.9-1.4~$\mu$m wavelength range, the model spectra get bluer as metallicity decreases while all other parameters are held fixed. In detail, the predicted F110W$-$$J$ colors are 0.83 mag for [m/H] = 0, 0.43 mag for [m/H] = $-$0.5 dex, and $-$0.35 mag for [m/H] = $-1$ dex. Thus, if WISEA 1534$-$1043 had $T_{\rm{eff}} = 500$ K (the lowest available temperature with sufficient metallicity sampling in the \citealt{Tremblin_2015} set of models), our F110W$-$$J$ synthetic color analysis would suggest $-1 \lesssim $ [m/H] $\lesssim -0.5$ for WISEA 1534$-$1043. We note that, when the results of HST observing program 16243 (PI: Marocco) are published, measured F110W magnitudes for roughly a dozen more T/Y dwarfs will become available\footnote{\url{https://www.stsci.edu/hst/phase2-public/16243.pro}}.

From Figure \ref{fig:reduced_pm}, it can be seen that WISEA 1534$-$1043, if a Y dwarf, would have the sixth most extreme $J-$W2 color among all Y dwarfs, suggesting that it may be very cold even in comparison to the broader population of Y dwarfs\footnote{This comparison is made based on measured central values of $J-$W2 per object, not accounting for the uncertainties on $J-$W2 color.}. The five known/suspected Y dwarfs with redder $J-$W2 color than WISEA 1534$-$1043 are: CWISEP J040235.55$-$265145.4 \citep[$T_{\rm{eff}} \approx$ 370~K;][]{Meisner_2020a, Leggett_2021}, WD 0806$-$661B \citep[$T_{\rm{eff}} = 377 \pm 88$~K;][]{Luhman_2011, Kirkpatrick_2021a}, WISEA 0830+2837 \citep[$T_{\rm{eff}} \sim 300$-350~K;][]{Bardalez_2020}, WISE 0855$-$0714 \citep[$T_{\rm{eff}} \approx 250$~K;][]{Luhman_2014}, and WISE 182831.08+2650 \citep[$T_{\rm{eff}}$ $\sim$ 325 K);][]{Cushing_2021b}.

\subsection{The T/Y Boundary Phototype of Ross 19B}
\label{sec:ross19B}

\begin{figure*}
\begin{center}
\includegraphics[width=6.0in]{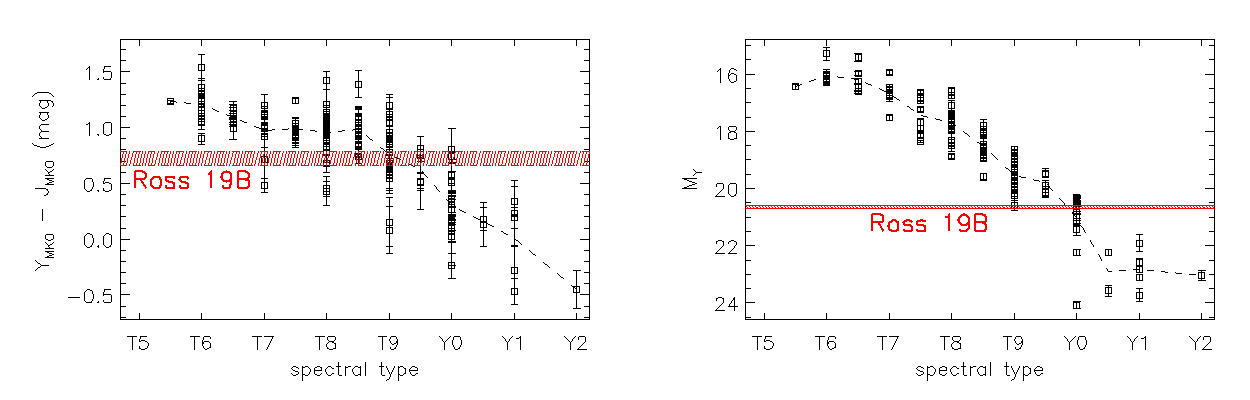}
\caption{Our new Gemini $Y$-band detection of Ross 19B allows us to further place it in context with the population of late T and early Y dwarfs. Left: $Y_{MKO}$$-$$J_{MKO}$ color as a function of spectral type, for the population of known late T and Y dwarfs (black squares with vertical error bars). The $Y_{MKO}$$-$$J_{MKO}$ measurement for Ross 19B is shown as a hatched red rectangle. Right: $Y$-band absolute magnitude versus spectral type for late T and Y dwarfs, plus our Ross 19B measurement in red. The plotted $M_Y$ value for Ross 19B assumes that Ross 19B has a parallax equal to that of Ross 19 \citep[available from Gaia;][]{GAIA}. In both panels the dashed black line shows the mean trend for known objects as a function of spectral type. The $Y_{MKO}$$-$$J_{MKO}$ color and $M_Y$ measurements for Ross 19B favor a spectral type between T9 and Y0. \label{fig:ross19B}}
\end{center}
\end{figure*}

Figure \ref{fig:ross19B} contextualizes Ross 19B with respect to the populations of late-T and Y dwarfs in ways newly enabled by our Gemini $Y$-band detection of this object. The photometric data remain consistent with Ross 19B lying near the T/Y boundary, favoring a spectral type between T9 and Y0. This is true both in terms of the $Y$$-$$J$ trend as a function of spectral type (left panel of Figure \ref{fig:ross19B}) and the $M_Y$ trend as a function of spectral type (right panel of Figure \ref{fig:ross19B}). $H$-band photometry for Ross 19B would help to distinguish between late T and early Y\footnote{At the time that we selected Ross 19B for observations as part of our Gemini program, the $H$-band filter was not available for use on GNIRS.}, as the blue side of $H$-band bears the signature onset of NH$_{3}$ absorption characteristic of Y dwarfs but not T dwarfs \citep[e.g.,][]{Cushing_2011}. If Ross 19B is confirmed to be a Y dwarf, it would be the widest known Y-dwarf companion to either a main sequence star or white dwarf.

\section{Conclusion}
\label{sec:conclusion}

Using the Gemini North and Gemini South observatories, we have provided crucial, previously lacking near-infrared photometry for a population of old and cold brown dwarfs thought to include the first known esdTs and perhaps the first known Y-type subdwarf. Our photometry and subsequent analyses have highlighted CWISE 0738$-$6643 and CWISE 2217$-$1454 as relatively extreme in terms of kinematics and $J$$-$W2 color (a proxy for temperature) among the esdT candidate population. These two objects represent the most promising `bridges' between WISEA 1534$-$1043 and the broader esdT candidate sample. Completing the JHK photometry for CWISE 0738$-$6643 and CWISE 2217$-$1454 by obtaining a $K$-band detection of CWISE 2217$-$1454 should be prioritized, and would bear on the question of whether metal-poor T dwarfs have red or rather blue $J$$-$$K$ colors. NIR spectroscopy of both CWISE 0738$-$6643 and CWISE 2217$-$1454 should be prioritized, so as to best measure their true metallicities via detailed model comparisons.

Continued searches for more examples of very cold and metal poor brown dwarfs remain vital. It seems unlikely that WISEA 1534$-$1043 would be in truth so anomalous and disjoint from the rest of the T/Y dwarf population as it appears today. Pinpointing a new set of examples that span the present gap between WISEA 1534$-$1043 and other known substellar objects would help us to establish a more clear mapping between the physical and observational properties of the lowest luminosity brown dwarfs. Further modeling of low-temperature giant exoplanet and brown dwarf atmospheres, down to effective temperatures $<$ 500 K and metallicities $< -1$ dex, is also greatly needed.

\vspace{5mm}
\facilities{Gemini North,
            Gemini South,
            WISE,
            NEOWISE,
            $Spitzer$}

\software{Astropy \citep{astropy:2013, astropy:2018},
          DRAGONS \citep{Labrie_2019},
          IDLUTILS,
          WiseView \citep{Caselden_2018}}

\section*{acknowledgments}
We thank Nicolas Lodieu for sharing the \cite{Lodieu_2022} spectrum of WISEA 1810$-$1010 with us. We thank Kevin Luhman for sharing the NIR spectrum of WISE 2105$-$6235 with us. Based on observations obtained at the international Gemini Observatory, a program of NSF's NOIRLab, which is managed by the Association of Universities for Research in Astronomy (AURA) under a cooperative agreement with the National Science Foundation on behalf of the Gemini Observatory partnership: the National Science Foundation (United States), National Research Council (Canada), Agencia Nacional de Investigacion y Desarrollo (Chile), Ministerio de Ciencia, Tecnologia e Innovacion (Argentina), Ministerio da Ciencia, Tecnologia, Inovacoes e Comunicacoes (Brazil), and Korea Astronomy and Space Science Institute (Republic of Korea). This work was enabled by observations made from the Gemini North telescope, located within the Maunakea Science Reserve and adjacent to the summit of Maunakea. We are grateful for the privilege of observing the Universe from a place that is unique in both its astronomical quality and its cultural significance. This publication makes use of data products from the \textit{Wide-field Infrared Survey Explorer}, which is a joint project of the University of California, Los Angeles, and the Jet Propulsion Laboratory/California Institute of Technology, funded by the National Aeronautics and Space Administration. The UHS is a partnership between the UK STFC, The University of Hawaii, The University of Arizona, Lockheed Martin and NASA. The VISTA Data Flow System pipeline processing and science archive are described in \cite{Irwin_04}, \cite{Hambly_2008} and \cite{Cross_2012}. This research has made use of the NASA/IPAC Infrared Science Archive, which is funded by the National Aeronautics and Space Administration and operated by the California Institute of Technology.  This research has benefitted from the Y Dwarf Compendium maintained by Michael Cushing at \url{https://sites.google.com/view/ydwarfcompendium/}.

\appendix

\begin{figure*}
\begin{center}
\plotone{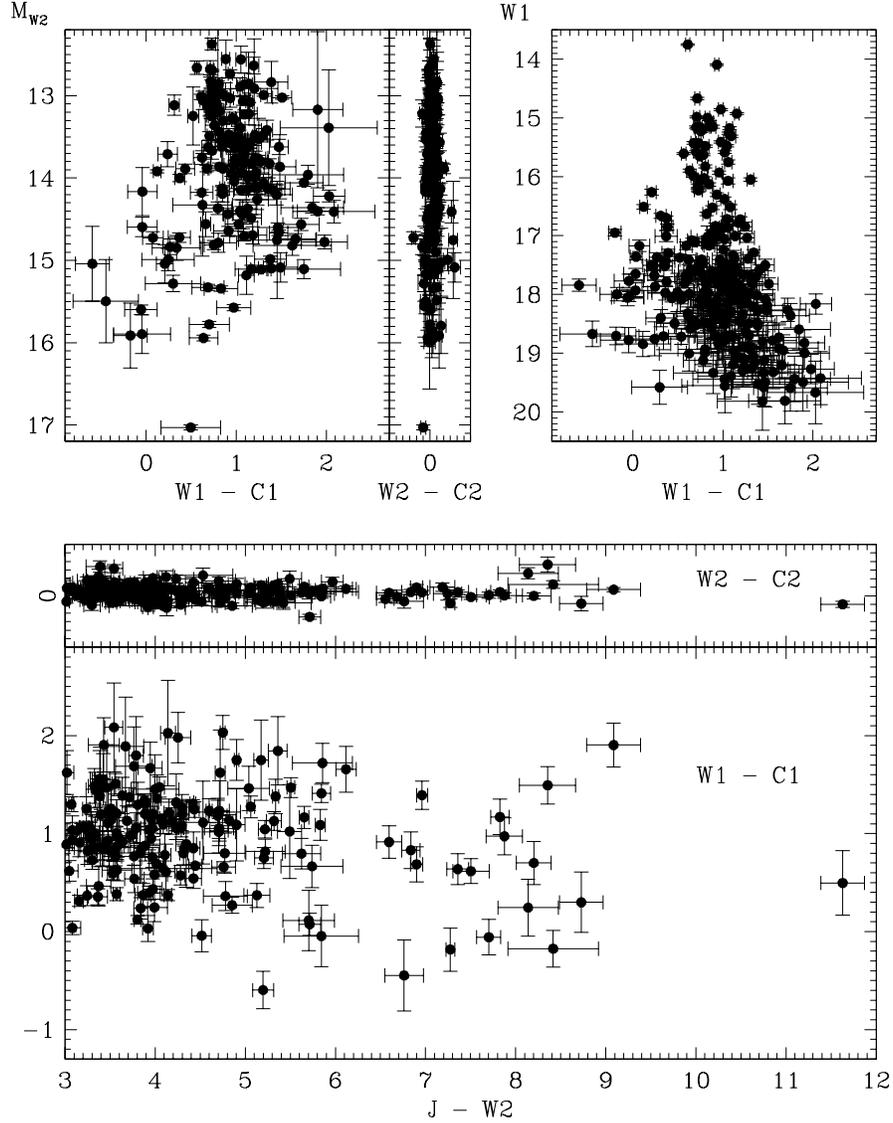}
\caption{Comparison of WISE W1/W2 and \textit{Spitzer} ch1/ch2 photometry for mid-late T dwarfs, Y dwarfs, and T/Y subdwarf candidates. W2 and ch2 remain tightly correlated with effectively zero systematic offset across the range of $M_{W2}$ and $J$$-$W2 values spanned by this sample. On the other hand there is considerable scatter between W1 and ch1 magnitudes, with this scatter becoming larger for increasing $M_{W2}$, W1, and $J$$-$W2. This suggests that caution ought to be exercised when attempting to select cold, low-metallicity candidates based on WISE W1$-$W2 color, as has been previously attempted/suggested in \cite{Meisner_2021} and \cite{Brooks_2022}.
\label{fig:wise_spitzer}}
\end{center}
\end{figure*}

In the course of this work, we took care to update our photometry tables for mid/late T, Y,  and metal-poor T/Y dwarf candidates with CatWISE2020 photometry \citep{Marocco_2021}, which we found to be superior to foregoing data products such as AllWISE \citep{Cutri_2013} and CatWISE Preliminary \citep{Eisenhardt_2020}. Figure \ref{fig:wise_spitzer} provides a comparison of CatWISE2020 and \textit{Spitzer} photometry for a large sample of late-T and Y dwarfs with the relevant \textit{Spitzer} and WISE photometry available. Figure \ref{fig:wise_spitzer} shows that W1$-$ch1 (ch1 is abbreviated as ``C1'' in the plot annotations) becomes quite noisy for very faint/cold/red populations, specifically as $M_{W2}$, W1, and/or $J$$-$W2 increase. This suggests that those attempting to select metal-poor brown dwarf candidates should exercise caution to avoid over-interpreting what may appear to be anomalously blue W1$-$W2 color. The large scatter for W1 is presumably due to lesser sensitivity in WISE as compared to \textit{Spitzer}, plus the frequent WISE blending that ensues due to its $\sim 6\farcs5$ FWHM PSF (compared to $\approx 1\farcs7$-$2\farcs0$ FWHM for \textit{Spitzer}/IRAC ch1 and ch2).

\begin{figure*}
\begin{center}
\plotone{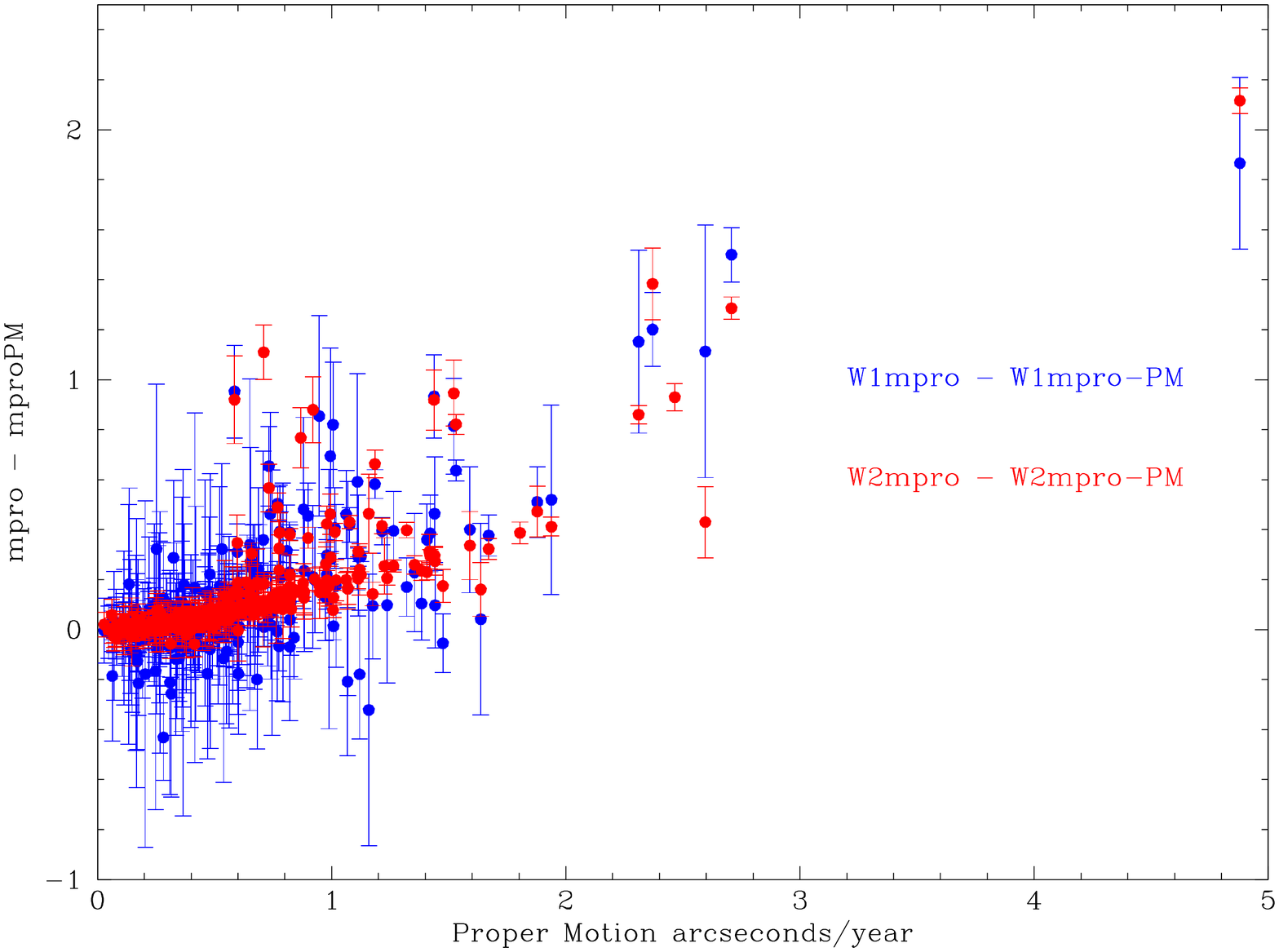}
\caption{Comparison of CatWISE2020 magnitudes fitted with (without) accounting for proper motion, \texttt{w?mpro\_pm} (\texttt{w?mpro}), based on the sample of T and Y dwarfs shown in Figure \ref{fig:color_color_sandy}. As the total proper motion (horizontal axis value) gets larger, a greater fraction of flux is ``missed'' by \texttt{w?mpro}, resulting in \texttt{w?mpro} magnitudes that are systematically too faint. We therefore caution that it is important to use \texttt{w1mpro\_pm} and \texttt{w2mpro\_pm} rather than \texttt{w1mpro} and \texttt{w2mpro} for objects with total motions $\gtrsim 1''$~yr$^{-1}$, even though \texttt{w1mpro\_pm} and \texttt{w2mpro\_pm} are not the default CatWISE2020 fluxes provided via IRSA.
\label{fig:mpro_pm}}
\end{center}
\end{figure*}

Figure \ref{fig:mpro_pm} illustrates the importance of using CatWISE magnitudes that account for proper motion, \texttt{w1mpro\_pm} and \texttt{w2mpro\_pm}, especially for objects with $\mu \gtrsim 1''$~yr$^{-1}$. By $\mu \sim 1''$~yr$^{-1}$, the WISE photometric bias incurred by neglecting to use the \texttt{\_pm} columns reaches $\approx 20$+\%.

\bibliography{m_poor_BDs_2022.bib}{}
\bibliographystyle{aasjournal}

\end{document}